\documentclass[twocolumn,twoside]{article} \usepackage {amsmath,amsthm,bm,cmap}
\usepackage[unicode]{hyperref}\usepackage[margin=1in,centering]{geometry}
\newcommand\aut {Leonid A.~Levin}\newcommand\ttl {Climbing LP Algorithms}
 \usepackage[utf8]{inputenc}\usepackage[T1]{fontenc}\usepackage{microtype}
 
\begin{document}\small\frenchspacing

 \newcommand\hreff[1] {{\footnotesize\href{https://#1}{https://#1}}}
 \newcommand\p[1]{\paragraph{#1}}
 \newcommand\edf{{\raisebox{-3pt}{$\,\stackrel{\text{\tiny df}}{=}\,$}}}
 \newcommand\Id {{\mathbf I}} \newcommand\s\sigma
 \newcommand\ov\overline \newcommand\un\underline
 \newcommand\1{{\un1}} \newcommand\0{{\mathbf{\ov 0}}}
 \newcommand\trm[1]{{\bf\em #1}} \newcommand\get{\leftarrow}
 \newcommand\ue{{\un e}} \newcommand\ce{{\ov e}}
 \newcommand\tri{{\mbox{\LARGE$\mathbf\triangleleft$}}}
 \renewcommand\v{{\mathbf v}} \renewcommand\d\delta \renewcommand\l\lambda

 \title {\vspace*{-2pc}\ttl} \date{} \author{\aut\\ Boston University\thanks
 {Computer Science dept., 111 Cummington Mall, Boston, MA 02215;
 Home page: \hreff{www.cs.bu.edu/fac/Lnd}}}\maketitle

\begin{abstract}NP (search) problems allow easy correctness tests for
solutions. {\em Climbing} algorithms allow also easy assessment of how close
to yielding the correct answer is the configuration at any stage of their
run. This offers a great flexibility, as how sensible is any deviation from
 the standard procedures can be instantly assessed.\par An example is
the \trm {Dual Matrix Algorithm (DMA)} for linear programming, variations of
which were considered by A.Y.Levin in 1965 and by Yamnitsky and myself in 1982.
It has little sensitivity to numerical errors and to the number of
inequalities. It offers substantial flexibility and, thus, potential for
further developments. \end{abstract}

\section {Introduction}

A \trm {Climbing} algorithm $A(x)$ is supplied with an easily computable \trm
{valuation} $V(s)$. $A$ runs by iterating a \trm {step} transformation $T(s)$
starting from $s\get x$ and ending with output $y\get s$ once the required
valuation $V(s)=R(x)$ is reached. The efficiency of the algorithm reflects the
effort required to compute $T,V$ and the progress in $V$ toward $R(x)$ each
iteration of $T$ assures.

The following account is an example of such algorithm. It adds algebraic
details to previously published geometric versions. Some of these details
exploit the flexibility provided by its climbing nature to help with worst
case-performance by deviating in appropriate cases from the standard procedure.
The main point is to emphasize the flexibility assured by this climbing nature
as an example that may be useful to follow for some other algorithms and
problems.

\subsection {The Idea}

\p{Notations.} $\edf$ indicates definitions. $\det(C)$ denotes the
determinant of a matrix $C$; the Euclidean norm of a vector $b$ is $|b|$.
For clarity row vectors may be underlined: $\1=(1,\ldots,1)$, column overlined:
$\0=(0,\ldots,0)$. $\Id$ is the identity matrix, $\ue_k,\ce_k$ are its $k$-th
row and column. We are to solve a system of inequalities $Ax>\0$ for a rational
vector $x$, given an $m{\times}n$ integer matrix $A$. Our inequalities are
$a_kx>0$ for $a_k\edf\ue_kA$, linearly independent for $k\le n$. Let $n$
 and all entries of $A$ be $<l$ bits long, so each $a_k$ has $<L=nl$ bits.
 By Hahn-Banach Theorem, the system $Ax>\0$ is inconsistent iff $bA=\un0$
for some vector $\un b\ge\un0\ne b$. The same holds if $|bA|<b\ov1/4^L$.

\pagebreak\p{The DMA}\hspace{-6pt} searches for $b$ in the form $dB$, where
matrix $B$ has no negative entries, $C\edf BA{=}V^{-1}$, $d\edf u
V{=}\1D{>}\un0$ for a diagonal $D$, $u \edf\sum_{k\le n}a_k$. We must grow
$b\ov1$ to~${>}4^L$. This growth is hard to keep monotone, so a lower bound
$\log(n!\det(DC)){<}n\log(b\ov1){+}3L$ is grown instead. It is ${-}\log$ of the
{\bf volume of the simplex} $\tri^B$ with faces $Cx{=}\0$, $u x{=}1$, vertices
$VD^{-1}\ce_k$, $\0$, and center $\v{=}VD^{-1} \ov1/(n{+}1)$. The original
simplex $\tri^o$ starts with $B^o_{k,k}=1$, $B^o_{k,k'\ne k}=0$.

It turns out that by incrementing a single entry of $B$ one can always increase
$\ln\det(DC)$ by $>1/2n^2$, as long as $x=\v$ fails the $Ax>\0$ requirement.
This provides an $O(n^3L)$ steps algorithm. Each step takes $O(n^2)$ arithmetic
operations, on $O(L)$-bit numbers and one call of a procedure which points an
inequality $a_k x>0$ violated by a given solution candidate $x$. This call is
the only operation that may depend on the number $m$ of inequalities, which
could even be an infinite family with an oracle providing the violated $a_k$.

\subsection {Some Comparisons}

The above bound has $n$ times more steps than Ellipsoid Method (EM). However,
the EM is much more demanding with respect to the precision with which the
numbers are to be kept. The simplex $\tri^B$ cannot possibly fail to include
all solutions of $Ax>\0$, $u x<1$, {\em whatever} $B$ with no negative
entries is taken. In contrast, the faithful transformation of ellipsoids in the
EM is the only guarantee that they include all solutions.

Also, for $m=O(n)$ several Karmarkar-type algorithms have lower polynomial
complexity bounds. Yet, they work in the dual space and their bounds are in
terms of the number $m$ of inequalities, while the above DMA bound is in terms
of $n$. For DMA, $m$ may even be infinite, e.g., forming a ball instead of a
polyhedron. Then dual-space complexity bounds break down, while the DMA
complexity is not affected (as long as a simple procedure finds a violated
inequality for any candidate $x$).

To assure fast progress, numbers are kept with $O(L)$ digits. This bound cannot
be improved since some consistent systems have no solutions with shorter
entries. Yet, this or any other precision is not actually {\em required} by
DMA. Any rounding (or, indeed, any other deviation from the procedure) can be
made any time as long as $\log\det(DC)$ keeps growing, which is immediately
observable. This leaves DMA open to a great variety of further developments. In
contrast, an inappropriate rounding in the EM, can yield ellipsoids which,
while still shrinking fast, lose any relation to the set of solutions and
produce a false negative output.

\subsection {A Historical Background}

The bound $\det(DC)$ is inversely proportional to the volume of $\tri^B$, which
parallels the EM. Interestingly, in history this parallel went in the opposite
direction: The simplexes enclosing the solutions were first used by A.Y.Levin
in 1965 \cite{al} and the EM was developed by Nemirovsky and Yudin in 1976
\cite{yn} as their easier to analyze variation.

The A.Y.Levin's algorithm starts with a large simplex guaranteed to contain all
solutions. Its center of gravity is checked, and, if it fails some inequality,
the corresponding hyperplane cuts out a ``half'' of the simplex. The process
repeats with the resulting polyhedron. Each cut decreases the volume by a
constant factor and so, after some number $q(n)$ of steps the remaining body
can be re-enclosed in a new smaller simplex. Only a weak upper bound
$q(n)<n\log n$ was proven by A.Y.Levin; it did not preclude the simplex from
turning into a too complex to manipulate in polynomial time polyhedron.

Nemirovsky and Yudin replaced simplexes with ellipsoids and made $q(n)=1$. Both
they and A.Y.Levin used real numbers and looked for approximate solutions with
a given accuracy. Khachian in 1979 \cite{kh} modified the EM for rationals and
exact solutions. Yamnitsky and myself in 1982 \cite{yl} proved $q(n)=1$ for the
original A.Y.Levin's simplex splitting method. Below, an algebraic version of
that geometric algorithm and some implementation improvements are
considered and analyzed.

\section {Main Algorithm and Analysis}

Let $u{\edf}\sum_{k\le n}a_k$, $C{\edf}BA{=}V^{-1}$, $d{\edf}u V$,
 $d_k{\edf}d\ce_k$,\\ $c_k{\edf}\ue_kC$, $v_k\edf V\ce_k/d_k$.\\
 For some $i,j,s$ let $a{\edf}a_i$, $v{\edf}v_j$, $t\edf(s^2{-}1)av$. 
 Then $C'\edf(B{+}\ce_j\ue_i/td_j)A{=}(V')^{-1}$,
 $d'{\edf}u V'$, $\d_k{\edf}d'\ce_k/d_k$.

With $\s\edf\Id{+}\ce_jaV/td_j$, $C'{=}\s C$,
 $\det(\s){=}(1{+}av/t)=\\1{+}1/(s^2{-}1)$, $V{=}V'\s$, $d{=}d'\s$.\hfil
 Thus, $d_k{=}d'\s\ce_k=d_k\d_k{+}\d_jav_kd_k/t$ and 
$1{=}\d_k{+}\d_jav_k/t{=}\d_k{+}\frac{\d_jav_k}{(s^2{-}1)av}$.\\
 Taking $k{=}j$, $1{=}\d_j(1{+}1/(s^2{-}1))$, $\d_j{=}(s^2{-}1)/s^2=\\1{-}1/s^2
=1/\det(\s)$, $\d_k{=}1{-}\frac{av_k}{s^2av}$. Our gain is $\ln\l$ for
 $\l\edf\prod_k\d_k\det(C')/\det(C)=\prod_k\d_k\det(\s){=}\prod_{k\ne j}\d_k$.

 Now $\v\edf\sum_kv_k$ take $i,j$ with $a_i\v\le0$, $a_iv_j{=}\max_ka_iv_k$. 
 Then $\d_k\ge\d_j$, $\sum_k\d_k=n-\frac{a\v}{s^2av}\ge n$, and $\prod_{k\ne j}
 \d_k\ge \d_j^{(n-2)}(\d_j{+}n(1{-}\d_j))=(1{-}1/s^2)^{n-1}(1{+}n/(s^2{-}1))$.

 So, $\ln\l\ge(n{-}1)\ln(1{-}1/s^2){+}\ln(1{+}n/(s^2{-}1))$. 
 For $s=n{-}1$ and $f(s)\edf s\ln(1{+}1/s)$ this is $f(s){-}f(s{-}1)>1/2n^2$.

 This $>1/2n^2$ gain holds if $s$ is accurate to $O(l)$ digits, 
 so $t$ can be rounded to $O(l)$ significant digits, too.

\vfil\pagebreak\section {Some Improvements}

\p {Inverting matrices} may take cubic time, but when a matrix with
a known inverse is moderately modified, Sherman-Morrison formula gives
its inverse in $O(n^2)$ steps. In our case the inverse of $C'$ is
$V'\edf V{-}\frac{vaV}{avs^2}$. Finally, the following {\bf occasional
deviations from the standard step} help the worst-case performance and
also illustrate the potential allowed by the flexibility of the algorithm.

\p {Digits.} The nodes $v^o_k$ of the starting simplex $\tri^o$ lie in a $4^L$
ball. Rounding $t$, $DB$ to $O(L)$ digits preserves the $>1/2n^2$ gain in steps
with $\max_k\log|v_k|<4L$. Yet, at some steps a longest edge $(v_i,v_j)$ of
$\tri^B$ may grow up to $2^{O(nL)}$ long. But there would be only $O(n)$ of such
 steps, since they allow large gains in $\ln\det(DC)$ as follows. Let $\un w\edf
(v_j{-}v_i)^T\!\!$, $M\edf\max_kwv^o_k$, $m\edf\max_{k',k}w(v^o_k{-}v^o_{k'})$,\\
 $t'\edf\,(wv_j{-}M)/m$, $t\edf\,\max\{0,t'{-}1\}/(|w|^2d_i)$,\\
 $h_{i,j}\edf(Mu-w)t$. $\tri^B$ has area $p$ of its projection along $w$ and
 volume $p|w|/n$. Its slice of height $m/|w|{=}O(4^L)$ cut by $h_{i,j}x{\ge}0$,
 $h_{j,i}x{\ge}0$ encloses $\tri^o$. Note that $h_{i,j}=bA$ for $b\edf
 h_{i,j}V^oB^o=t(M\1{-}wV^o)B^o\ge\un0$.
 The new simplex replaces faces $c_i,c_j$ in
$\tri^B$ with $c_i{+}h_{i,j}$, $c_j{+}h_{j,i}$. 
It is up to $3$ times wider and higher than the above slice. 
So its volume is $O(4^L3^np)$. This gains $>\log|w|-3L$ in $\log\det DC$.

\p{Maintaining Sparsity.} Given a row $\un b$ of $B$ let $S\edf\{i:b_i{\ne}0\}$.
 When $|S|$ exceeds $2n$, we can (in $O(n^3)$ arithmetic operations)
simplify $B$ to get $|S|{\le}n$ without changing $C$.
We find a set $F$ of $|S|{-}n$ linearly independent vectors $\un f$ such
that $\{i:f_i{\ne}0\}\subset S$ and $fA{=}\un0$.

Then repeat the following. Use an $f{\in}F$ and $i{\in}S$ with maximal
$\frac{|f_i|}{b_i}$ to annul $b_i$ via $b\get b-\frac{b_i}{f_i}f$. This
preserves $bA$, keeps $b{\ge}\un0$, and shrinks $S$. As $S$ looses an entry $i$
we use an $f{\in}F$ to annul the $i$-th component in all other vectors in $F$
and drop $f$ from $F$.

\vfil\end{document}